\documentclass[letterpaper,oneside,english]{amsart}
\usepackage{times}
\usepackage[T1]{fontenc}
\usepackage[latin9]{inputenc}
\usepackage{amsmath}
\usepackage{graphicx}
\usepackage{amssymb}
\usepackage{esint}
\makeatletter

\newtheorem{definitn}{Definition}
\newtheorem{lemma}{Lemma}
\newenvironment{lyxlist}[1]
{\begin{list}{}
{\settowidth{\labelwidth}{#1}
 \setlength{\leftmargin}{\labelwidth}
 \addtolength{\leftmargin}{\labelsep}
 }}
{\end{list}}
\newtheorem{thm}{Theorem}

\makeatother

\usepackage{babel}

\begin{document}
\renewcommand{\baselinestretch}{0.97}
 \pagestyle{empty}
\title{The Transport Capacity of a Wireless Network is a  Subadditive Euclidean
Functional}

\author{Radha Krishna Ganti and Martin Haenggi\\
Department of Electrical Engineering\\
University of Notre Dame\\
Indiana-46556, USA\\
\{rganti,mhaenggi\}@nd.edu}
\maketitle
\thispagestyle{empty}

\begin{abstract}
The transport capacity of a dense ad hoc network with $n$ nodes
scales like $\sqrt{n}$. We show that the transport capacity divided
by $\sqrt{n}$ approaches a non-random limit with probability one
when the nodes are i.i.d. distributed on the unit square. We prove
that the transport capacity under the protocol model is a subadditive
Euclidean functional and use the machinery of subadditive functions
in the spirit of Steele to show the existence of the limit. 
\end{abstract}

\section{Introduction}
\footnotetext[1]{ 1-4244-2575-4/08/\$20.00  \copyright  2008 IEEE}
Consider a wireless network of $n$ nodes in a unit square on the
plane. Finding the capacity region of this setup is an unsolved problem. Transport
capacity is a metric which, in a loose sense, indicates the sum rate
of the network while incorporating the notion of distance. It was
shown in \cite{gupta2000cwn} and \cite{franceschetti2004cgc} that the transport capacity (TC) is
  $\Theta(\sqrt{n})$. More precisely when\emph{
no cooperative communication techniques are used} (except for pure
relaying of   packet), the transport capacity $T$ is   bounded by
\cite{gupta2000cwn,franceschetti2004cgc} \[
C_{2}\sqrt{n}<T(X_{n})<C_{1}\sqrt{n}\]
when $X_{n}=\left\{ x_{1},\cdots,x_{n}\right\} $ are $n$ nodes uniformly
distributed on the unit square and $n$ is large. The lower bound is
provided by Franceschetti etal. using percolation theory. When cooperative
communication techniques are used, the transport capacity scales like
$n$ \cite{ozgur2007hca}. When one restricts the network to act like
a packet network without any cooperative techniques (except packet
relaying), TC exhibits a nice geometric behavior. While it has been
proved that TC scales like $\sqrt{n}$, the question whether the limit
\begin{equation}
\lim_{n\rightarrow\infty}\frac{T(X_{n})}{\sqrt{n}}\label{eq:main}\end{equation}
exists remained open when the $n$ nodes $x_{i},\ 1\leq i\leq n$
are i.i.d distributed in a unit square. In this paper we show that
\eqref{eq:main} converges to a constant with probability one.  This technique can be easily extended
to show that \[
\lim_{n\rightarrow\infty}T(X_{n})/n^{(d-1)/d}=A_{d}\ \text{a.s.}\]
 when the nodes $x_{i}$ are distributed i.i.d in $[0,1]^{d},\ d\geq2$
and $A_{d}$ is a constant depending only on the system parameters
and the dimension $d$. We show that transport capacity has a geometric
flavor similar to the minimum spanning trees (MST), Euclidean matching
(EM) problem and Euclidean travelling salesman problem (TSP). The existence of a limit is more of
a mathematical interest, but the techniques used in proving the limit
will help in a better understanding of scheduling and routing mechanisms.

The paper is organized as follows. In Section \ref{sec:System-Model},
we introduce the communication model and the definition of TC. In
Section \ref{sec:Limit-Theorems}, we present the geometrical properties
of TC and derive the limit. In Theorem \ref{thm:main} we prove the
convergence result when the nodes are i.i.d uniformly distributed
on a unit square. Theorem \ref{thm:non-uniform} provides a similar
result when the nodes are i.i.d distributed with a general PDF $f(x)$.

\section{\label{sec:System-Model}System Model }

We assume the protocol model \cite{gupta2000cwn} for communication between
two nodes, i.e., a node located at $x_{i}$ can communicate successfully
to a node located at $x_{j}$ if the ball centered around $x_{j}$
with radius $\beta|x_{i}-x_{j}|$, $\beta>1$, does not contain any other
transmitter. When the communication is successful, we assume one packet
of information is transmitted%
\footnote{Basically we are neglecting noise. Neglecting noise can make the achievable
rate unbounded. So we cap the link capacity to unity. Alternatively
we can assume a packet of information transmitted.%
}

\begin{definitn}
\label{def:Transport-Capacity:-For}Transport Capacity: For $n$ nodes
$\left\{ x_{1},x_{2},\cdots,x_{n}\right\} \subset \mathbb{R}^{2}$, the
transport capacity of these $n$ nodes is defined as \[
T\left(\left\{ x_{1},x_{2},\cdots,x_{n}\right\} \right)=\sup_{\mathcal{S}}\left[\sum_{(i,j)\in[1,2..n]^{2}}\lambda_{ij}|x_{i}-x_{j}|\right]\]
where the supremum is taken over the supportable rate pairs $\mathcal{S}$.
The set $\mathcal{S}$ can also be thought of as the set of all scheduling
and routing algorithms. The set $\mathcal{S}$ contains scheduling
algorithm with fixed source and destination pairs. $\lambda_{ij}$
denotes the information rate that node $x_{i}$ can communicate to
$x_{j}$ (we don't count the relaying nodes). Observe that \emph{the
definition of $T(\left\{ x_{1},\cdots,x_{n}\right\} )$ depends only
on the location of the nodes $x_{i},1\leq i\leq n$.} We make the
following assumptions:
\end{definitn}
\begin{enumerate}
\item Time is discretized.
\item Message set for each source destination pair is independent\emph{.}
\item \emph{No cooperative communication techniques are used.} 
\item $T\left(\{x_{1}\}\right)=0$
\end{enumerate}
We will consider two cases. One with no constraint on $\lambda_{ij}$
and the other with the following constraint.

\emph{Constraint} 1: $\lambda_{ij}>0$ for some $j$ for every $i$,
i.e., $\max_{j}\lambda_{ij}>0$, $\forall i$ \\
\\
\emph{Notation}: Let $B(x,r)$ denote a ball of radius $r$ centered
around $x$. For a set $A$, the complement is denoted by the set
$A^{c}$. For a finite set $A$, $|A|$ denotes the cardinality of
the set $A$. We will use $(A\rightarrow B)$ to denote the set of
transmissions with transmitters in $A$ and receivers in $B$.

\section{\label{sec:Limit-Theorems}Limit Theorems}

In this section we show the existence of the limit \eqref{eq:main} using
tools from subadditive sequences. A sequence $\left\{ a_{m}\right\} $
is subadditive if $a_{m+n}\leq a_{m}+a_{n}$. By a theorem of Fekete,
we have that $\lim a_{m}/m=\inf\left(a_{m}/m\right)$ exists. Similar
results hold when the sequence is superadditive. Most of the geometrical
quantities like the length of a minimum spanning tree on $n$ points,
or a Euclidean matching of $n$ points are not strictly subadditive.
They have a small correction factor, i.e., of the form $a_{m+n}\leq a_{m}+a_{n}+c(m,n)$.
If the growth of $c(m,n)$ can be controlled, the existence of the limit
can be proved. When the underlying sequences are random variables,
the existence of the limit is provided by a classical result of Kingman \cite{kingman1968ets}.
Steele   has used such a frame work to prove
the existence of the limit of a weakly subadditive sequences in the
geometrical setting \cite{steele1981sef}. The geometrical quantities which exhibit such
subadditivity are coined {}``Subadditivie Euclidean functionals''.
We will use the framework of Steele to prove the existence of the
limit \eqref{eq:main}. For doing so, we first establish the weak
subadditivity of TC and other required properties. We start by  introducing
the following bound on TC which was proved in \cite{kumar-book}.
We state it for convenience.

\begin{lemma}
\label{lem:[Sphere-packing-bound]}{[}Sphere packing bound] The transport
capacity of $n$ nodes $\left\{ x_{1},x_{2},\cdots,x_{n}\right\} $
located in a square $[0,t]^{2}$ is bounded by $Ct\sqrt{n}$, where
$C$ is a constant not depending on the location of nodes or $n$.
\end{lemma}
\begin{proof}
See Section 2.5 in \cite{kumar-book}
\end{proof}

\subsection{Basic properties of TC}

In this subsection, unless indicated, $X_{n}=\left\{ x_{1},x_{2},\cdots,x_{n}\right\} $
are deterministic points on the plane. From the definition of $T$,
we can consider $T$ as a functional on finite subsets of $\mathbb{R}^{2}$.
We then have 

\begin{lyxlist}{00.00.0000}
\item [{\emph{(A0)}}] $T\left(X_{n}\right)$ is a continuous function of
$\left\{ x_{1},x_{2},\cdots,x_{n}\right\} $ and hence measurable.
\item [{\emph{(A1)}}] $T\left(aX_{n}\right)$ $=aT\left(X_{n}\right)$
for all $a>0$.
\item [{\emph{(A2)}}] $T\left(X_{n}+x\right)=T\left(X_{n}\right)$ for
all $x\in\mathbb{R}^{2}$ where $X_{n}+x=\left\{ x_{1}+x,x_{2}+x,\cdots,x_{n}+x\right\} $ 
\end{lyxlist}
(A1) and (A2) imply $T$ is a Euclidean functional.

\begin{lyxlist}{00.00.0000}
\item [{\emph{(A3)}}] \emph{Monotone property}: $T\left(X_{n}\cup \{x\}\right)\geq T\left(X_{n}\right)$.
\emph{The above monotone relation does not hold true with constraint
$1$. }
\item [{\emph{(A4)}}] Finite variance: \[
\text{Var }T\left(\left\{ x_{1},x_{2},\cdots,x_{n}\right\} \right)<\infty\]
when $x_{i}$ are independently and uniformly distributed on $[0,1]$.
This follows from Lemma \ref{lem:[Sphere-packing-bound]}.\\

\end{lyxlist}
The next lemma provides an estimate, which is used to bound the correction
factor in the subadditivity of TC.

\begin{lemma}
\label{pro:critical}  
Consider the scenario in which   nodes in a square $S=[0,t]^{2}\subset\mathbb{R}^{2}$ can only be  transmitters
 that have to communicate with   receivers outside the square $S$ in
a single hop. If we restrict the maximum Tx-Rx distance to be $c_{1}t$,
then the transport capacity in this setup is upper bounded by $c_{2}t$.
\end{lemma}
\begin{proof}
For a transmitter receiver pair $(x_{k},y_{k})$ denote \[
D_{k}=\cup_{x\in\text{line}(x_{k},y_{k})}B\left(x,\frac{(\beta-1)}{2}|x_{k}-y_{k}|\right)\]
 i.e., the $\frac{(\beta-1)}{2}|x_{k}-y_{k}|$ neighborhood of the
line joining $x_{k}$ and $y_{k}$ .%
\begin{figure}
\begin{centering}
\includegraphics[clip,width=2.2in]{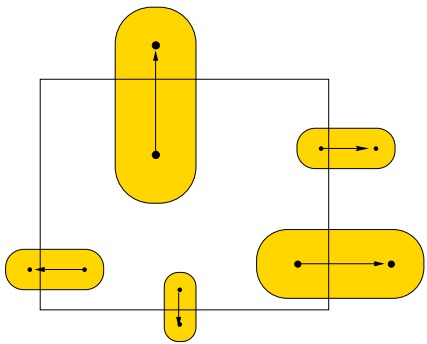}
\par\end{centering}

\caption{Illustration of the Proof. The coloured regions represent { $D_{k}$}}

\label{fig:square}
\end{figure}
 See Figure \ref{fig:square}. For all the successful Tx-Rx pairs,
the regions $D_{k}$ are disjoint. The proof of the above is identical
to Theorem 3.3 in \cite{kumar-book}. In our case we have that the
transmitters are inside the square $[0,t]^{2}$. Let the contending
transmitter-receiver distances be $\left\{ r_{1},r_{2},\cdots,r_{n}\right\} $.
Since the receivers are outside the box and each transmitter-receiver
pair cuts the boundary, we have \[
2\frac{\beta-1}{2}\left(r_{1}+r_{2}+\cdots+r_{n}\right)\leq4t\]
 Hence the single hop transport capacity in this case is upper bounded
by $4t/(\beta-1)$
\end{proof}

From the previous lemma we observe that the TC  is constrained by  the perimeter of the domain $A$ which contains the
nodes, when the transmissions are from the set $\left(A\rightarrow A^{c}\right)$.
In some sense this indicates that TC is maximized when the communication
is local, i.e., short hops. In the next lemma we prove that the bottleneck in a multihop network for achieving TC is the maximum packing 
of scheduling on a plane.  Loosely speaking
\emph{unconstrained TC metric is more suitable for a single-hop network.}

\begin{lemma}
\emph{\label{lem:flatten}Multihop to single-hop conversion }{[}\emph{Flattening
the network}]:\emph{ }Any scheme which achieves the TC consists of
only single hops, i.e., every packet reaches the destination from
source in a single hop. 
\end{lemma}
\begin{proof}
Suppose a flow $\lambda_{ij}$ is helped by $n$ nodes. Now instead
of assisting this flow, each of these $n$ nodes send their own independent
packets for a single hop they serve. By simple triangle inequality
this procedure guarantees a single hop scheme that achieves the same
or larger TC.
\end{proof}

In the next lemma we prove a form of subadditivity. We use the fact
that the network can be visualized of as a single-hop network and
the idea that the TC is maximized by local communications. See Figure
\ref{fig:illus_main}, for a graphical illustration of the proof.

\begin{lemma}
\label{lem:main}{[}\emph{Cutting Lemma}]: Consider a square $A=[0,t]^{2}\subset\mathbb{R}^{2}$
and let $X=\{x_{1}\ldots x_{k}\}\subset A$ denote a set of $k$ nodes.
Divide $A$ into $m^{2}$ squares of equal sides with length $t/m$
and denote each square by $A_{i}$. We then have \[
T(X)\leq\sum_{i=1}^{m^{2}}T(X\cap A_{i})+Cmt\]

\end{lemma}
\begin{proof}
Let some scheme achieve the TC of $X$. By Lemma  \ref{lem:flatten} the scheme that achieves TC is a single hop scheme.
We now focus on a single square $A_{i}$. There are three types of
transmissions, $(A_{i}\rightarrow A_{i})$, $(A_{i}\rightarrow A_{i}^{c})$
and $(A_{i}^{c}\rightarrow A_{i})$. %
\begin{figure}
\begin{centering}
\includegraphics[width=2.2in]{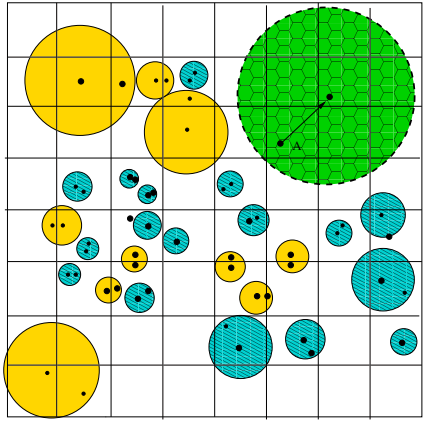}
\par\end{centering}

\caption{Proof technique: The blue hashed circles (dark hashed) correspond
to { $(A_{i}\rightarrow A_{i})$} and the TC contribution can be bounded
by { $T(A_{i})$}. The yellow unhashed circles corresponds to { $(A_{i}\rightarrow A_{i}^{c})$}.
However these cannot contribute much to the TC by Lemma \ref{pro:critical}.
The maximum contribution from them is { $cm^{2}t/m=cmt$}. There
is a trade-off between { $(A_{i}\rightarrow A_{i}^{c})$} and the large
transmissions denoted by hashed yellow region on the top corner. Observe
that when the Tx-Rx distance is greater than { $a=2\sqrt{2}t/(m(\beta-1))$},
there can be a maximum of one transmission per square (as in the green
comb circle on the right corner).}
\label{fig:illus_main}
\end{figure}
See Figure \ref{fig:illus_main}. The contribution of transmissions
from $A_{i}$ into $A_{i}$ to the TC, can be upper bounded by $T(X\cap A_{i})$.
Hence the total contribution by $(A_{i}\rightarrow A_{i}$), $1\leq i\leq m^{2}$
is upper bounded%
\footnote{This is true since we consider $T(X\cap A_{i})$ as only a function
of $X\cap A_{i}$. %
} by $\sum_{i=1}^{m^{2}}T(X\cap A_{i})$. The only transmissions which
involve $A_{i}$, to be accounted are $(A_{i}\rightarrow A_{i}^{c})$
and $(A_{i}^{c}\rightarrow A_{i})$.   Denote the contribution
of these transmissions to the TC by $\tilde{T}$. Let $F(A_{k})$
denote the set of feasible transmitters in square $A_{k}$ with receivers
in $A_{k}^{c}$. By the sphere packing bound we have \[
\sum_{k=1}^{m^{2}}\sum_{(x,y)\in F(A_{k})}|x-y|^{2}\leq Ct^{2}\]
Let $b_{k}=\sum_{(x,y)\in F(A_{k})}|x-y|$. So we require to bound
$\tilde{T}=\sup\left\{ \sum_{k=1}^{m^{2}}b_{k}\right\} $ where the
supremum is taken over all the feasible transmissions. Let the number
of squares with all of their transmission distance less than $a=2\sqrt{2}t/(m(\beta-1))$
be $\eta$. Denote this set of squares by $C_{a}\subset\left\{ 1,\cdots,m^{2}\right\} $.
So we have $|C_{a}|=\eta$ and $\tilde{T}=\sup\left\{ \sum_{k\in C_{a}}b_{k}+\sum_{k\in C_{a}^{c}}b_{k}\right\} $.
Let $A_{k}\in C_{a}^{c}$. We then have $|F(A_{k})|=1$. Hence $\sum_{k\in C_{a}^{c}}b_{k}$
is upper bounded by (since the maximum number of transmitters is $m^{2}-\eta$)

\[
c_{1}t\sqrt{m^{2}-\eta}\]
 For the other set $C_{a}$ with Tx-Rx distances less than $a$, by
Proposition \ref{pro:critical}, the contribution $\sum_{k\in C_{a}}b_{k}$
to the transport capacity is upper bounded by \[
c_{2}\frac{t}{m}\eta\]
So we have \[
\tilde{T}\leq c_{1}t\sqrt{m^{2}-\eta}+c_{2}\frac{t}{m}\eta,\ 0\leq \eta\leq m^{2}\]
 The maximum value of the right hand side for the given range of $\eta$
is $ctm$.
\end{proof}
\begin{thm}
Let $\left\{ Q_{i}:1\leq i\leq m^{2}\right\} $ be a partition of
the square $[0,1]^{2}$ into squares with edges parallel to the axis
and length $m^{-1}$. Let $tQ_{i}=\{x;x=ty,y\in Q_{i}\}$. 

\emph{(A5)} Subadditivity: Let $X=\left\{ x_{1},x_{2}\cdots x_{n}\right\} $.
We then have \begin{equation}
T(X\cap[0,t]^{2})\leq\sum_{i=1}^{m^{2}}T(X\cap tQ_{i})+Ctm\label{eq:subadd}\end{equation}

\end{thm}
\begin{proof}
This follows immediately from Lemma \ref{lem:main}.
\end{proof}
Equation \eqref{eq:subadd}, does not imply subadditivity, but only
a weaker form of it. Nevertheless it is denoted  as subadditive property
for convenience.

\begin{thm}
\label{thm:main}Let $x_{i},1\leq i\leq n$, and $x_{i}$ are i.i.d
uniformly distributed in $[0,1]^{2}$. If $\lambda_{ij}$ is not constrained
then \begin{equation}
\lim_{n\rightarrow\infty}\frac{T\left(\left\{ x_{1},x_{2},\cdots,x_{n}\right\} \right)}{\sqrt{n}}=A_{2}\label{eq:main_convergence}\end{equation}
with probability one. $A_{2}$ is a constant depending only on $\beta$.
\end{thm}
\begin{proof}
The conditions (A1) to (A5) indicate that $T$ is a monotone, Euclidean
functional with finite variance and satisfies subadditivity. \eqref{eq:main_convergence}
follows from the subadditive Euclidean convergence theorem by Michael
Steele \cite[Thm 1]{steele1981sef}.
\end{proof}
Observe that in the above theorem, monotonicity of $T$ is necessary.
Hence  it does not hold with constraints on $\lambda_{ij}$,
i.e., constraint 1.\emph{ }To overcome this we require to prove the
\emph{smoothness} of $T$.

Let $Q_{i},i\in\{1,2,3,4\}$ be a partition of the unit square into
$4$ equal squares. By Theorem \ref{thm:main} we have

\begin{lyxlist}{00.00.0000}
\item [{\emph{(A6)}}] \[
T(F)\leq\sum_{i=1}^{4}T(F\cap Q_{i})+C\]
\end{lyxlist}
where $F$ is any finite set in $[0,1]^{2}$. 
The above result follows from (A5) but we numbered it for convenience.
In the next Lemma we prove the smoothness of $T(A)$ with respect
to the cardinality of $A$. Observe that this sense of continuity
is different from (A0).

\begin{lemma}
\textbf{\label{lem:(A7)-Continuity:-For}}\emph{(A7)} {[}Smoothness]:
For finite point sets $F,G \subset[0,1]^{2}$  (observe $F$ and
$G$ need not be disjoint), we have\begin{equation}
|T(F\cup G)-T(G)|<c\sqrt{|F|}\label{eq:continuity}\end{equation}
where $c$ is a constant that does not depend on $F$ and $G$.
\end{lemma}
\begin{proof}
We use the same trick as we did in Theorem \ref{lem:main}. We flatten
the network of $F\cup G$. The transmissions can be partitioned into
$(G\rightarrow G)$, $(F\rightarrow F)$, $(G\rightarrow F)$, $(F\rightarrow G)$.
The contribution of the transmissions $(G\rightarrow G)$ to TC can
be upper bounded by $T(G)$. Observe that the maximum cardinality
of the remaining transmissions can be $|F|$. So we have \begin{eqnarray*}
T(F\cup G) & < & T(G)+c\sqrt{|F|}\end{eqnarray*}

\emph{If we do not assume any constraint on $\lambda_{ij}$, then
we are done by the monotonicity. }If Constraint 1 has to be satisfied,
we  n have to prove \begin{eqnarray*}
T(F\cup G) & \geq & T(G)-c\sqrt{|F|}\end{eqnarray*}
We use time sharing to prove this. By Lemma \ref{lem:[Sphere-packing-bound]}, we have 
$T(F)<c_{1}\sqrt{|F|}$. So we can assume $T(G)>T(F)$ (otherwise there
is nothing to be proved). We use time sharing between the set of nodes,
$G$ and $F$. By time sharing the constraint that each node transmits
some data of its own is satisfied. So we obtain a transport capacity
of \begin{eqnarray}
&&\lambda T(G)+(1-\lambda)T(F)\label{eq:777}
 \end{eqnarray}
Choose \begin{eqnarray*}
1-\lambda & = & \frac{1}{\frac{T(G)}{T(F)}-1}\end{eqnarray*}
So if $T(G)>2T(F)$, we have $(1-\lambda)<1$ and \begin{eqnarray*}
 &  & T(G)-(T(G)-T(F))(1-\lambda)\nonumber \\
 & = & T(G)-T(F)\label{eq:888}\end{eqnarray*}

Otherwise we have $0<T(G)-T(F)\leq T(F)$. So from \eqref{eq:777},
we have \begin{eqnarray*}
 &  & T(G)-(T(G)-T(F))(1-\lambda)\nonumber \\
 & \geq & T(G)-T(F)(1-\lambda)\label{eq:999}\\
 & \geq & T(G)-T(F)\end{eqnarray*}
i.e., any time sharing will give a transport capacity greater than
$T(G)-T(F)$. So by time sharing we have constructed a scheme which
obeys constraint 1 and has a TC of at least $T(G)-T(F)$. Since $T(F\cup G)$
is the supremum over all such schemes we have, \begin{eqnarray*}
T(F\cup G) & \geq & T(G)-T(F)\\
 & \stackrel{(a)}{\geq} & T(G)-c\sqrt{|F|}\end{eqnarray*}
where $(a)$ follows from the sphere packing bound on the set $F$.
\end{proof}
\emph{(B-1)} We also have the following. Let $F$ and $G$ be any
finite subsets of $[0,1]^{2}$. Then \begin{eqnarray*}
 &  & |T(F)-T(G)|\\
 & \stackrel{(a)}{\leq} & |T(F)-T(F\cap G)|+|T(G)-T(F\cap G)|\\
 & \stackrel{(b)}{\leq} & c\left\{ \sqrt{|F\setminus(F\cap G)|}+\sqrt{|F\setminus(F\cap G)|}\right\} \\
 & \leq & \sqrt{2}c\left\{ \sqrt{|F\setminus(F\cap G)|+|G\setminus(F\cap G)|}\right\} \\
 & = & \sqrt{2}c\sqrt{|F\triangle G|}\end{eqnarray*}
where $(a)$ follows from triangle inequality and $(b)$ follows from
Lemma \ref{lem:(A7)-Continuity:-For}.

We now use the theorem from Rhee \cite {rhee1993mpa} to prove the existence of the 
limit when Condition 1 is satisfied.
From the conditions (A1) to (A8) we have the following convergence
of the mean and concentration around the mean. This result holds even
with Constraint 1 unlike Theorem \ref{thm:main}.

\begin{lemma}
\label{lem:uniform rate}Let $X_{n}=\left\{ x_{1},x_{2},\cdots,x_{n}\right\} $
denote $n$ i.i.d nodes in $[0,1]^{2}$. For the transport capacity
we have that \[
\lim_{n\rightarrow\infty}\frac{\mathbb{E}T\left(X_{n}\right)}{\sqrt{n}}=A_{2}\]
and \begin{eqnarray}
\mathbb{P}\left(\left|T(X_{n})-\mathbb{E}T(X_{n})\right|\geq t\right) & \leq & C\exp\left(-C_{1}\frac{t^{4}}{n}\right)\label{eq:conc}\end{eqnarray}

\end{lemma}
\begin{proof}
Follows from \cite[Thm 1]{rhee1993mpa}. Here we do not require monotonicity
and the complete subadditive hypothesis. Conditions (A6) and (A7) replace
those two. 
\end{proof}
If we choose $t$ to be $t\sqrt{n}$, we have the right hand side
of \eqref{eq:conc} is $\exp(-C_{1}t^{4}n)$. Equation \eqref{eq:conc}
also implies complete convergence i.e., for all $\epsilon>0$\[
\sum_{n>1}\mathbb{P}\left(\left|\frac{T(x_{n})}{\sqrt{n}}-A_{2}\right|>\epsilon\right)<\infty\]

\subsection{Non uniform  distribution of nodes}
In the previous subsection, we have proved the  existence of the limit when the nodes are uniformly distributed on an unit square. In this subsection we prove the existence of the limit and show its relation to $A_2$ when the nodes are distibuted with a PDF $f(x)$. In Lemma \ref{lem:main}, we proved an upperbound to $T(X_n)$ by the transport cpacity of disjoint subsets of $x_n$.  We now  prove a  lower bound to $T(X_n)$ by similar subsets of $X_n$. 
\begin{lemma}
\label{lem: glue2}{[}\emph{Asymptotic Glueing Lemma}] Consider two
bounded disjoint sets $A,B\subset\mathbb{R}^{2}$ and an infinite
sequence of nodes $\left\{ x_{i}\right\} $. Let $X_{n}=\left\{ x_{1},x_{2},\cdots,x_{n}\right\} $
be a subset of the sequence. We then have \begin{eqnarray}
 &  & T(X_{n}\cap A)+T(X_{n}\cap B)\label{eq:lemma_glue}\\
 & \leq & T(X_{n}\cap(A\cup B))+o(\sqrt{n})\end{eqnarray}
 
\end{lemma}
\begin{proof}
Consider the flattened networks of $A$ and $B$ which achieve the
TC of $A$ and $B$ respectively. Wlog we can assume we can assume
$T(A)=\Theta(\sqrt{n})$ and $T(B)=\Theta(\sqrt{n})$ (otherwise there
is nothing to prove). We have to find a scheme such that \eqref{eq:lemma_glue}
is satisfied. Consider the following. At any time, neglect all transmissions
with transmitter receiver distance greater than $\sqrt{\log(n)/n}$.
\begin{figure}
\begin{centering}
\includegraphics[width=2.2in]{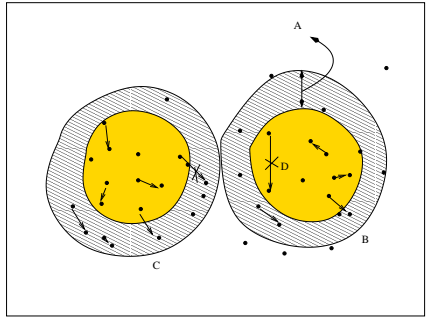}
\par\end{centering}

\caption{The hashed region is the boundary with thickness $2\sqrt{\log(n)/n}$.
We neglect all transmissions in the inside region with length greater
than $\sqrt{\log(n)/n}$.}

\label{fig:glue-lemma}
\end{figure}
The loss in TC by removing these transmissions is $\sqrt{n/\log(n)}$.
This is because, the loss is given by $\text{ }\max\left(\sum_{(i,j)\in\mathcal{T}}d_{ij}\right)$
with the following constraints

\begin{eqnarray*}
\begin{cases}
 & \sum_{(i,j)\in\mathcal{T}}d_{ij}^{2}<A\\
 & d_{ij}>\sqrt{\frac{\log(n)}{n}}\end{cases}\end{eqnarray*}
where $\mathcal{T}$ is the set of all feasible transmissions with
Tx-Rx distance greater than $\sqrt{n/\log(n)}$. The solution to the
above problem is $\sqrt{An/\log(n)}$. See Figure \ref{fig:glue-lemma}.
Now neglect all the nodes along the boundary of $A$ and $B$ in a
strip of width $2\sqrt{\log(n)/n}$. The maximum penalty because of
this is \begin{eqnarray*}
c\sqrt{\frac{\log(n)}{n}}\sqrt{n} & = & c\sqrt{\log(n)}\end{eqnarray*}
 Now operate $A$ and $B$ networks together except for the nodes
in the strip as mentioned above and the transmissions with Tx Rx lengths
greater than $\sqrt{\log(n)/n}$. So we are still left with a transport
capacity of (that can be achieved by the union). \begin{eqnarray*}
 &  & T(A)+T(B)-c\sqrt{\frac{n}{\log(n)}}-c_{2}\sqrt{\log(n)}\\
 & = & T(A)+T(B)-o(\sqrt{n})\end{eqnarray*}
 We can operate the neglected strips of $A$ and $B$, the neglected
transmissions and the others in a time sharing fashion with time shares
\[
\left(1-\frac{1}{n},\frac{1}{3n},\frac{1}{3n},\frac{1}{3n}\right)\]
 In the resulting network Constraint 1 is satisfied.
\end{proof}
We have the following lemma required to prove the limit when the nodes
are not uniformly distributed. We can generalize the previous Lemma
to $s$ disjoint squares to prove the following.

\begin{lemma}
\textbf{\label{lem:(A-9) }}\emph{(A-9)} Let $Q_{i},\ 1\leq i\leq s$
be a finite collection of disjoint squares with edges parallel to
the axes and let $x_{i}\in\mathbb{R}^{2},$ $1\leq i<\infty$ an infinite
sequence. Let $X_{n}=\left\{ x_{1},x_{2},\cdots,x_{n}\right\} $.
We then have \begin{eqnarray*}
\sum_{i=1}^{s}T\left(X_{n}\cap Q_{i}\right) & \leq & T\left(X_{n}\cap\cup_{i=1}^{s}Q_{i}\right)+o(\sqrt{n})\end{eqnarray*}
 
\end{lemma}
\begin{proof}
Follows from Lemma \ref{lem: glue2}.
\end{proof}
We now prove the limit theorem when the nodes are i.i.d. distributed
with a \emph{blocked distribution. }A blocked distribution is of the
form $\phi(x)=\sum_{i=1}^{s}1_{Q(i)}(x)$ where $Q(i)$ are disjoint
squares with edges parallel to the axes. We use the homogeneous property
of TC and the glueing lemma to prove the next lemma. Also observe that
$\phi(x)$ looks like a simple function. Extending the result to general
distributions is of more technical nature and is stated in Theorem
\ref{thm:non-uniform}.

\begin{lemma}
\label{lem:blocked}Let $Y_{i},\ 1<i\leq n$ be a sequence of i.i.d
random variables with bounded support and no singular part \cite{folland}.
Let the absolutely continuous part be given by $\phi(x)=\sum_{i=1}^{s}1_{Q(i)}(x)$
where $Q(i)$ are disjoint cubes with edges parallel to the axes.
Let $\mathcal{Y}_{n}=\left\{ Y_{1},\cdots,Y_{n}\right\} $ One then
has \[
\lim_{n\rightarrow\infty}\frac{T(\mathcal{Y}_{n})}{\sqrt{n}}=A_{2}\int_{R^{d}}\sqrt{\phi(x)}dx\]

\end{lemma}
\begin{proof}
We follow the method provided in \cite{steele1981sef}. Without loss
of generality, we assume that the support of RV $Y_{i}$ lies in $[0,1]^{2}$.
Since the $Q(i)$ are disjoint we have by Theorem \ref{thm:main},\begin{eqnarray}
T(\mathcal{Y}_{n}) & \leq & \sum_{i=1}^{s}T(\mathcal{Y}_{n}\cap Q(i))+Cs\label{eq:temp111}\end{eqnarray}
 We have that $\mathcal{Y}_{n}\cap Q(i)$ is uniform on $Q(i)$ except
for the un-normalized measure $m(Q(i))$. Using (A-1) and Theorem
\ref{thm:main}, we have

\begin{eqnarray*}
\lim_{n\rightarrow\infty} & \frac{T(\mathcal{Y}_{n}\cap Q(i))}{\sqrt{\sum_{j=1}^{n}1_{Q(i)}(y_{j})}} & =A_{2}\sqrt{m(Q(i))}\end{eqnarray*}
By the law of large numbers we have, \[
\sum_{j=1}^{n}1_{Q(i)}(y_{j})\sim n\alpha_{i}m(Q(i))\ a.s\]
 So \[
\lim_{n\rightarrow\infty}\frac{T(\mathcal{Y}_{n}\cap Q(i))}{\sqrt{n}}=A_{2}\sqrt{\alpha_{i}}m(Q(i))\]
 So using \eqref{eq:temp111}, we obtain \[
\lim\sup_{n\rightarrow\infty}\frac{T(\mathcal{Y}_{n})}{\sqrt{n}}\geq A_{2}\int\sqrt{\phi(x)}dx\]
 By Lemma \ref{lem:(A-9) }, 

\begin{equation}
T(\mathcal{Y}_{n})\geq\sum_{i=1}^{s}T(\mathcal{Y}_{n}\cap Q(i))+o(\sqrt{n})\label{eq:678}\end{equation}
 By using a similar procedure on \eqref{eq:678}, we have a similar
result on $\lim\inf$ and hence the lemma follows.
\end{proof}
The next theorem characterizes the limiting behavior of TC when the
nodes are not uniformly distributed.

\begin{thm}
\label{thm:non-uniform}Let $y_{i}$ be i.i.d random variables, with
PDF $f(x)$ (i.e., no singular part w.r.t Lebesgue measure) and bounded
support. We then have \[
\lim_{n\rightarrow\infty}\frac{T\left(y_{1},y_{2},\cdots,y_{n}\right)}{\sqrt{n}}=A_{2}\int_{\mathbb{R}^{2}}\sqrt{f(x)}dx\]
 
\end{thm}
\begin{proof}
Follows from (B-1), Lemma \ref{lem:blocked} and Theorem 3 in \cite{steele1988gre}
(Observe the above theorem is not proved when the measure of $y_{i}$
has singular support).
\end{proof}
We immediately observe that the constant $A_{2}\int_{\mathbb{R}^{2}}\sqrt{f(x)}dx$
is maximized when $y_{i}$ are uniformly distributed.

\section{Conclusion}

In this paper we have shown that the transport capacity of $n$ nodes
distributed i.i.d with bounded support, when scaled by $\sqrt{n}$
approaches a non-random limit. The existence of a limit is more of
a mathematical interest, but the techniques used in proving the limit
will help in a better understanding of scheduling and routing mechanisms.

\bibliographystyle{ieeetr}

\bibliography{point_process}

\end{document}